\documentclass{PoS}

\title{Measurement of $\Lambda_{\rm c}^{+}$ baryons and ${\rm D}_{\rm s}^{+}$ mesons in Pb--Pb collisions with ALICE}

\ShortTitle{$\Lambda_{\rm c}^{+}$ baryons and ${\rm D}_{\rm s}^{+}$ mesons in Pb--Pb collisions with ALICE}

\author{\speaker{Luuk Vermunt} for the ALICE Collaboration\\
        Institute for Subatomic Physics, Utrecht University\\
        E-mail: \email{luuk.vermunt@cern.ch}}

\abstract{
The study of heavy-flavour (charm and beauty) production is important to understand the properties of the Quark-Gluon Plasma (QGP) formed in ultra-relativistic heavy-ion collisions, since heavy quarks are produced in the initial stages of the collisions and subsequently interact with the medium throughout its evolution. In the QGP, strange quarks are expected to be abundantly produced and may recombine with charm quarks leading to an enhancement of the nuclear modification factor ($R_{\rm AA}$) of ${\rm D}_{\rm s}^{+}$ mesons compared to that of other charmed meson states at low and intermediate $p_{\rm T}$. The measurement of elliptic flow ($v_{2}$) of ${\rm D}_{\rm s}^{+}$ mesons is useful to determine the degree of thermalisation of charm quarks in the collective expansion of the QGP. In addition, charm quarks could recombine with light di-quark states in the medium, which would lead to an enhancement of the $\Lambda_{\rm c}^{+}/{\rm D}^{0}$ baryon-to-meson ratio as compared to that in pp collisions. Precise measurements of these particle species in Pb--Pb collisions, therefore, give a deeper insight into the hadronisation mechanisms that heavy quarks undergo in the strongly-interacting medium. In this contribution, the latest results measured with ALICE for ${\rm D}_{\rm s}^{+}$-meson and $\Lambda_{\rm c}^{+}$-baryon production in Pb--Pb collisions at $\sqrt{s_{\rm NN}}=5.02$~TeV, using data from Run 2 of the LHC are discussed.
}

\FullConference{
European Physical Society Conference on High Energy Physics - EPS-HEP2019 -\\
			10-17 July, 2019\\
			Ghent, Belgium}

\begin{document}

\section{Introduction}

Under extreme temperatures and/or densities, lattice QCD calculations predict a phase transition of nuclear matter to a colour-deconfined medium, the so-called Quark-Gluon Plasma (QGP)~\cite{Karsch:2006xs, Borsanyi:2010bp, Borsanyi:2013bia, Bazavov:2011nk}. Ultra-relativistic heavy-ion collisions provide suitable conditions for the QGP formation and for characterising its properties. Ideal probes are heavy quarks (charm and beauty), as they are predominantly produced in the early stage of the collisions in hard-scattering processes. Their formation time (approximate 0.1 and 0.03~${\rm fm}/c$ for charm and beauty~\cite{Andronic:2015wma}) is shorter than the production time of the QGP (between 0.3-1.5~${\rm fm}/c$~\cite{Liu:2012ax}) at LHC energies. In contrast, the thermal production and annihilation rates of charm and beauty quarks are expected to be negligible~\cite{BraunMunzinger:2007tn}.

In this contribution, recent measurement of strange open-charm meson, ${\rm D}_{\rm s}^{+}$, and open-charm baryon, $\Lambda_{\rm c}^{+}$, production by the ALICE Collaboration in the LHC Pb--Pb run of 2018 are discussed. These measurements provide unique insights into the charm quark hadronisation processes in the QGP, probing the interplay between the fragmentation and coalescence mechanisms. In the QGP, strange quarks are expected to be abundantly produced in the medium~\cite{Rafelski:1982pu, Koch:1986ud}, which may recombine with charm quarks. In addition, charm quarks could recombine with light di-quark states in the medium~\cite{Donoghue:1988ec, Sateesh:1991jt, Lee:2007wr}. If the coalescence mechanism plays an important role, as experimental data is suggesting for low and intermediate $p_{\rm T}$~\cite{Acharya:2018hre, Acharya:2018ckj}, an enhanced production of ${\rm D}_{\rm s}^{+}$ and $\Lambda_{\rm c}^{+}$ is expected compared to non-strange ${\rm D}$ mesons and their production in pp collisions.

The ${\rm D}_{\rm s}^{+}$-meson and $\Lambda_{\rm c}^{+}$-baryon production in Pb--Pb collisions at $\sqrt{s_{\rm NN}}=5.02$~TeV was measured by ALICE via exclusive reconstruction at mid-rapidity ($|y|<0.8$) in the hadronic decay channels $\rm D^{+}_{s} \to \phi \pi^{+} \to K^{+} K^{-} \pi^{+}$ ($c\tau \simeq$ 150~$\mu$m, ${\rm BR}=2.27\pm 0.08\%$) and $\rm \Lambda_{c}^{+} \to p K^{0}_{S} \to p \pi^{+} \pi^{-}$ ($c\tau \simeq$ 60~$\mu$m, ${\rm BR}=1.10\pm 0.21\%$)~\cite{Tanabashi:2018oca}. The raw yields were extracted via an invariant-mass analysis after having applied topological and particle-identification selections to enhance the signal-over-background ratio. These selections are performed independently for the different variables in the ${\rm D}_{\rm s}^{+}$-meson analysis, where for the $\Lambda_{\rm c}^{+}$-baryon analysis two BDT machine-learning algorithms were exploited. The AdaBoost and XGBoost machine-learning classifiers were used in parallel, averaging the fully corrected $\Lambda_{\rm c}^{+}$ $p_{\rm T}$-differential yields using a similar strategy as used in Ref.~\cite{Acharya:2017kfy}. Monte-Carlo simulations based on HIJING~\cite{Wang:1991hta} and PYTHIA6~\cite{Sjostrand:2006za} event generators and the GEANT3 transport package~\cite{Brun:1994aa} were used for the efficiency-times-acceptance corrections, as well as for the machine-learning training\footnote{Background candidates for the machine-learning training are taken from invariant-mass sideband regions in data.}. The fraction of prompt ${\rm D}_{\rm s}^{+}$ mesons and $\Lambda_{\rm c}^{+}$ baryons was estimated using a FONLL-based approach~\cite{Cacciari:1998it,Acharya:2017qps}. The centrality and the direction of the event plane were provided by the V0 scintillators. The measurement of the ${\rm D}_{\rm s}^{+}$ meson $v_{2}$ was performed with the scalar-product (SP) method~\cite{Voloshin:2008dg}.

Both analyses benefit from the tracking and particle identification capabilities of the ALICE central barrel detectors. A complete description of the ALICE apparatus and its performance can be found in Refs.~\cite{Aamodt:2008zz, Abelev:2014ffa}. The main detectors used in this analysis include the Inner Tracking System (ITS)~\cite{Aamodt:2010aa}, the Time Projection Chamber (TPC)~\cite{Alme:2010ke}, the Time-Of-Flight detector (TOF)~\cite{Akindinov:2013tea}, and the V0 detector~\cite{Abbas:2013taa} located inside a solenoidal magnet.

\section{Results}

\begin{figure}[!tb]
	\begin{center}
		\includegraphics[width=.45\textwidth]{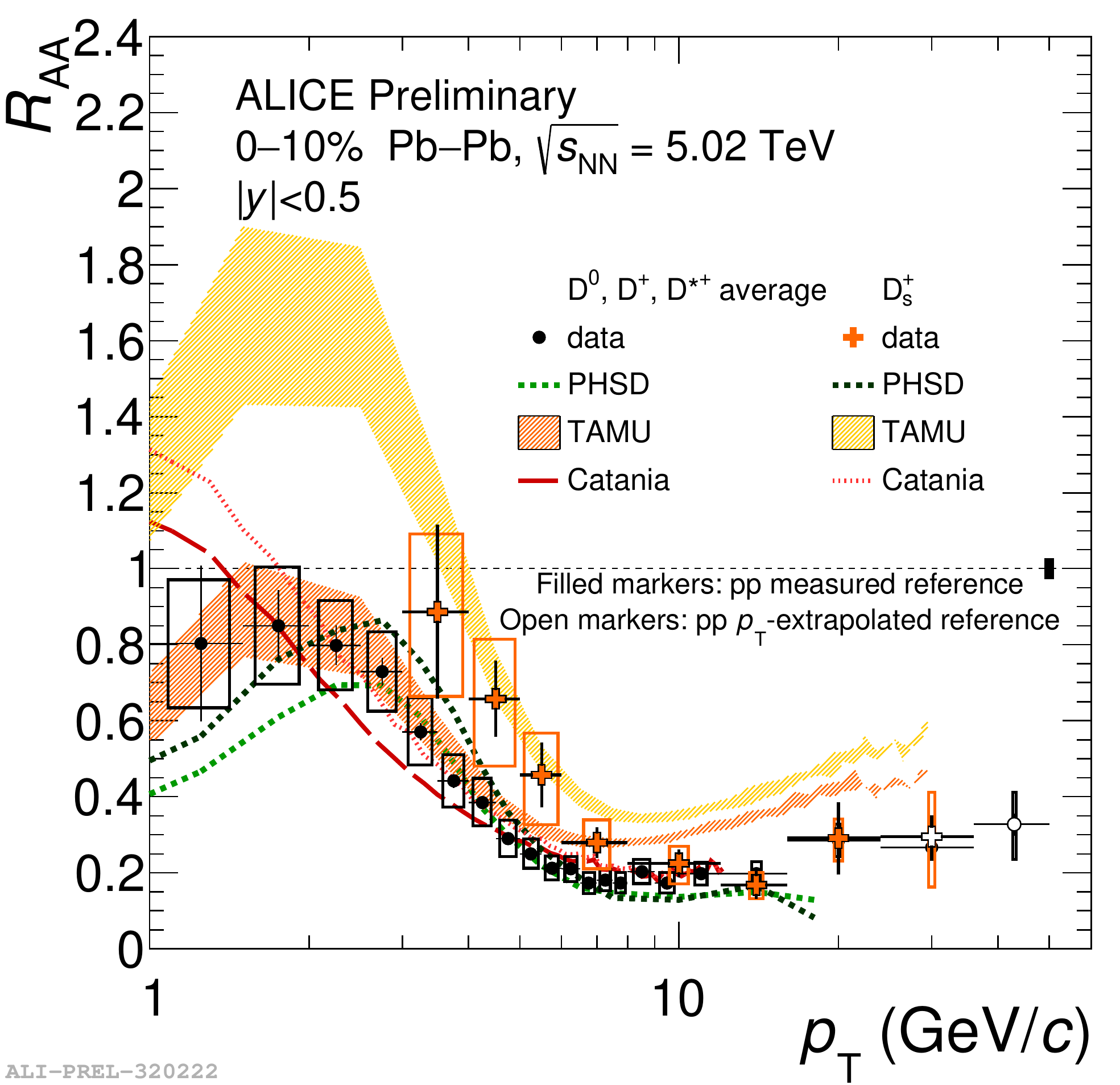}
		\hspace{0.04\textwidth}
		\includegraphics[width=.45\textwidth]{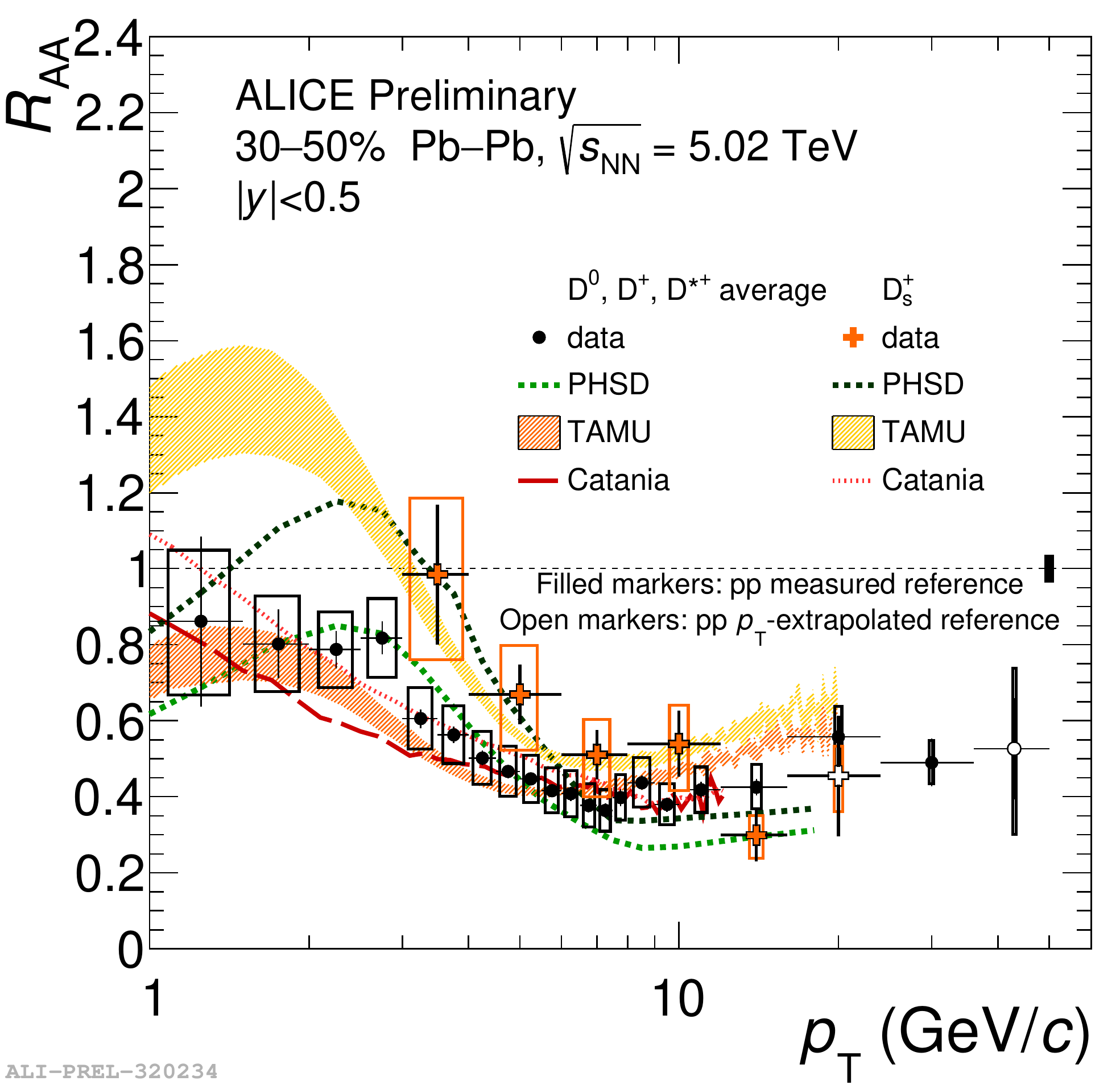}
		\caption{Average $R_{\rm AA}$ of ${\rm D}^{0}$, ${\rm D}^{+}$, and ${\rm D}^{*+}$ mesons and the $R_{\rm AA}$ of ${\rm D}_{\rm s}^{+}$ mesons in the 0-10\% (left) and 30-50\% (right) centrality class compared with the PHSD~\cite{Song:2015ykw}, TAMU~\cite{He:2014cla}, and Catania~\cite{Plumari:2017ntm} model calculations.}
		\label{fig:RaaDs}
	\end{center}
\end{figure}

Figure~\ref{fig:RaaDs} shows the nuclear modification factor, $R_{\rm AA}$, of prompt ${\rm D}_{\rm s}^{+}$ mesons measured in central (0-10\%) and semi-central (30-50\%) Pb--Pb collisions at $\sqrt{s_{\rm NN}}=5.02$~TeV. This observable is defined as the ratio between the $p_{\rm T}$-differential yield measured in nucleus-nucleus collisions (${\rm d}N_{\rm AA}/{\rm d}p_{\rm T}$) and the $p_{\rm T}$-differential production cross section in pp collisions (${\rm d}\sigma_{\rm pp}/{\rm d}p_{\rm T}$), scaled by the average nuclear overlap function $\langle T_{\rm AA} \kern-0.1em\rangle$. It is compared to the average $R_{\rm AA}$ of prompt non-strange ${\rm D}$ mesons. The $R_{\rm AA}$ of ${\rm D}_{\rm s}^{+}$ mesons is in general higher than that of non-strange ${\rm D}$ mesons, although the uncertainties are relatively large. The strange and non-strange ${\rm D}$-meson $R_{\rm AA}$ are compared to models that provide both observables. These three models, PHSD~\cite{Song:2015ykw}, TAMU~\cite{He:2014cla}, and Catania~\cite{Plumari:2017ntm}, all predict a similar increase in the common $p_{\rm T}$-range of the ${\rm D}_{\rm s}^{+}$-meson $R_{\rm AA}$ as seen in data. This increase is induced by hadronisation via coalescence in combination with the strangeness-rich QCD medium, taking also into account the different interaction cross sections in the hadronic phase of the system evolution.

\begin{figure}[!tb]
	\begin{center}
    \includegraphics[width=.443\textwidth]{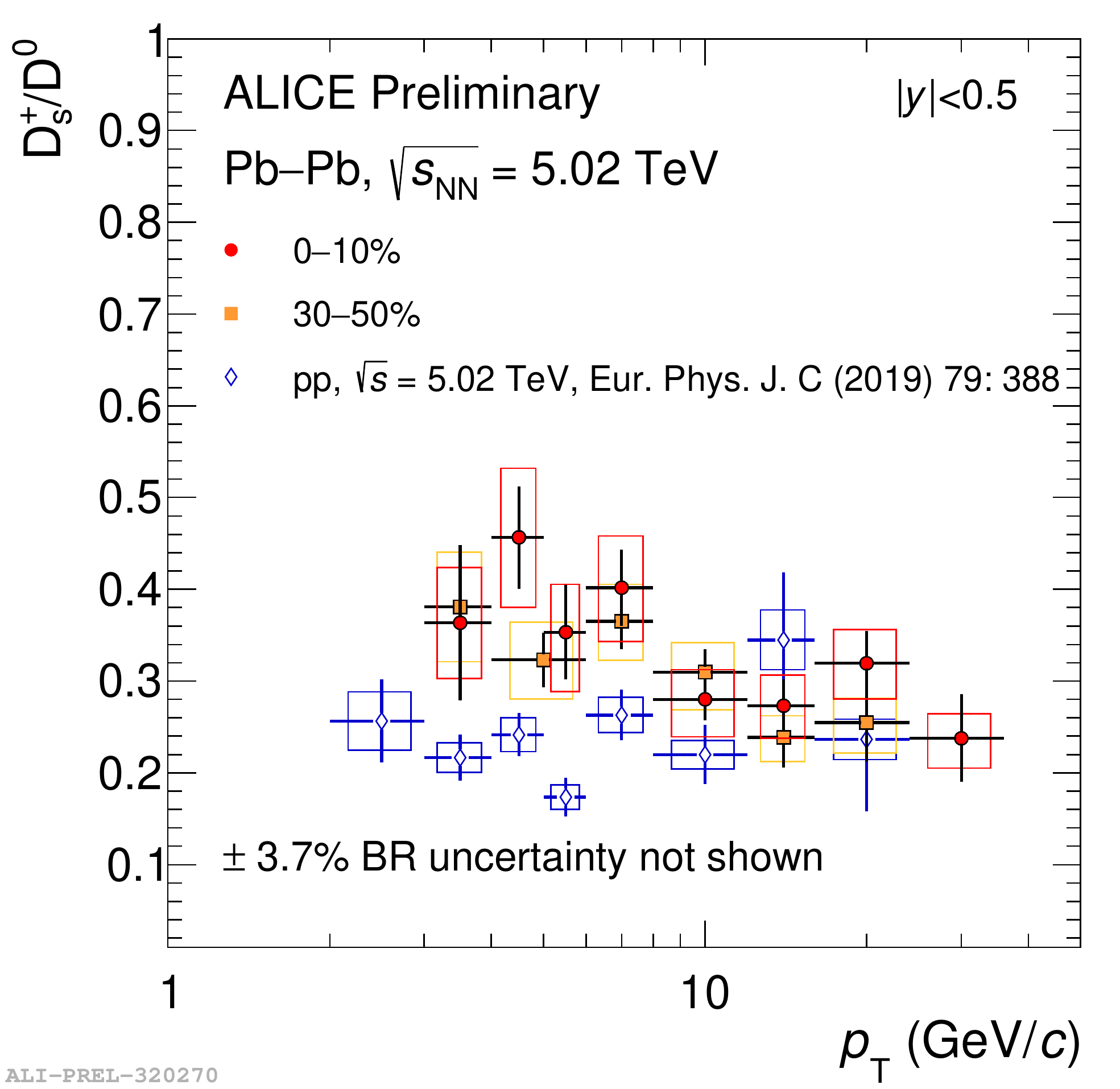}
		\hspace{0.04\textwidth}
    \includegraphics[width=.45\textwidth]{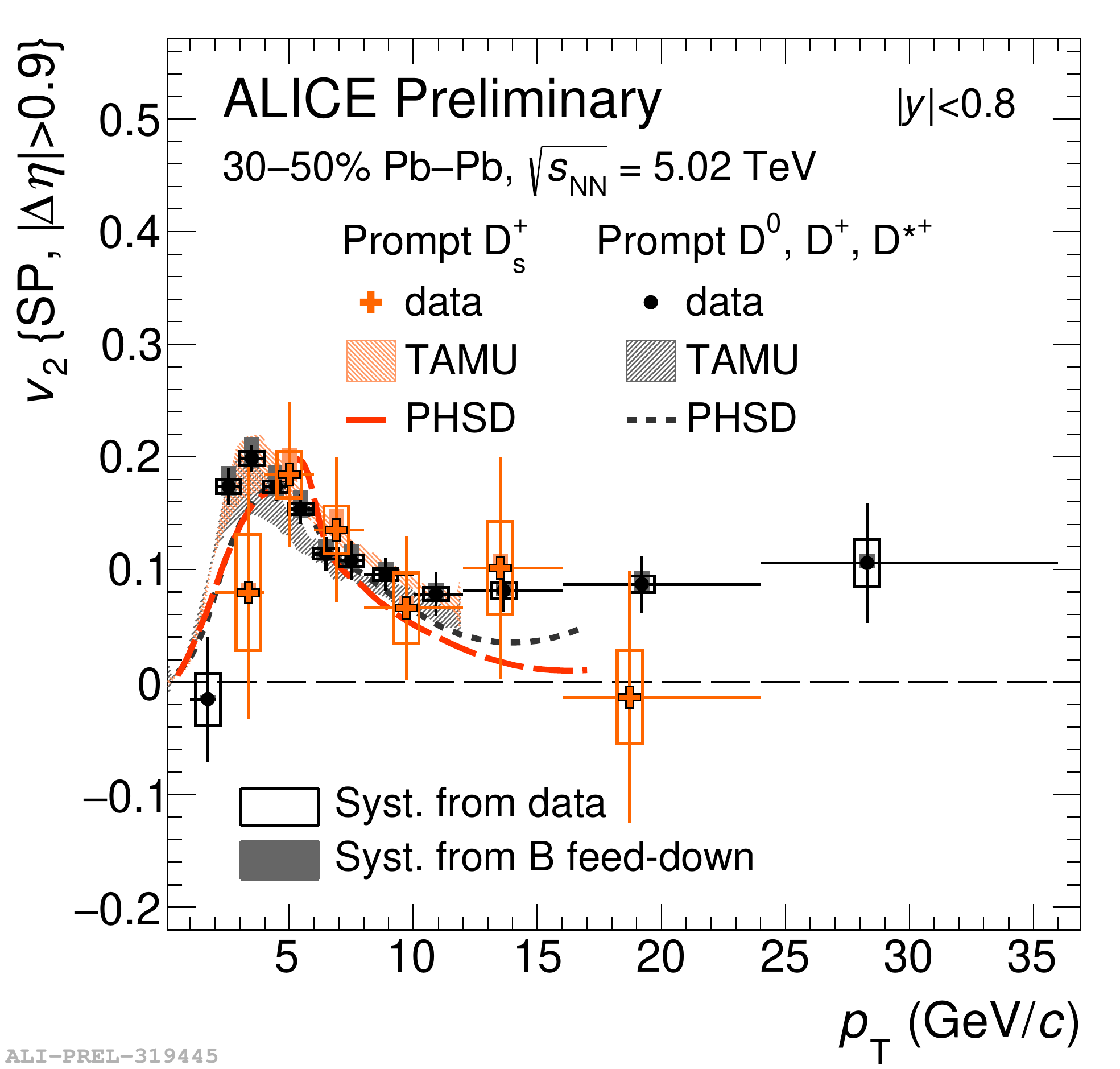}
		\caption{Left: The ${\rm D}_{\rm s}^{+} / {\rm D}^{0}$ ratio of prompt yields as function of $p_{\rm T}$ for pp and Pb--Pb collisions. Right: Average prompt ${\rm D}^{0}$, ${\rm D}^{+}$, and ${\rm D}^{*+}$ $v_{2}$ and the ${\rm D}_{\rm s}^{+}$-meson $v_{2}$ at $\sqrt{s_{\rm NN}}=5.02$~TeV in the 30-50\% centrality class, compared to the PHSD~\cite{Song:2015ykw} and TAMU~\cite{He:2014cla} model calculations.}
		\label{fig:v2DsandDsoverD0}
	\end{center}
\end{figure}

An alternative way to display the modification of the charm-quark hadronisation in the presence of a QGP is to compare the ratios between the measured yields of ${\rm D}_{\rm s}^{+}$ and ${\rm D}^{0}$ mesons in Pb--Pb and pp collisions (see left panel of Figure~\ref{fig:v2DsandDsoverD0}). A hint of a larger ratio up to $p_{\rm T} = 8$~GeV$/c$ in Pb--Pb than in pp collisions is found. This observation points to the same concept of hadronisation of charm quarks via coalescence in the QGP as for light quarks.

In the right panel of Figure~\ref{fig:v2DsandDsoverD0}, the prompt ${\rm D}_{\rm s}^{+}$-meson elliptic flow, $v_{2}$, in mid-central (30-50\%) Pb--Pb collisions is compared to that of non-strange ${\rm D}$-mesons and the PHSD~\cite{Song:2015ykw} and TAMU~\cite{He:2014cla} model predictions. The measurement of these azimuthal anisotropies, defined\footnote{In this definition, the $\phi$ is the particle-momentum azimuthal angle, $\Psi_{\rm n}$ is the symmetry-plane angle relative to the n$^{\rm th}$ harmonic, and the brackets denote the average over all the measured particles in the considered events.} as $v_{\rm n} = \langle \cos{n (\phi - \Psi_{\rm n})} \kern-0.1em\rangle$, are sensitive to the fraction hadronising via coalescence. In addition, the participation of the charm quark in the collective dynamics of the underlying medium and the degree of thermalisation in the medium can be probed at low $p_{\rm T}$, while at high $p_{\rm T}$ the $v_{2}$ is dominated by the path-length dependence of the partons' in-medium energy loss mechanisms. In the common $p_{\rm T}$-range, the measured $v_{2}$ is found to be compatible within uncertainties for ${\rm D}_{\rm s}^{+}$ mesons, non-strange ${\rm D}$ mesons, as well as with both theoretical calculations. With the current uncertainties it is not yet possible to conclude about the underlying processes that can cause a difference in the elliptic flow of strange and non-strange ${\rm D}$ mesons.

\begin{figure}[!tb]
	\begin{center}
		\includegraphics[width=.47\textwidth]{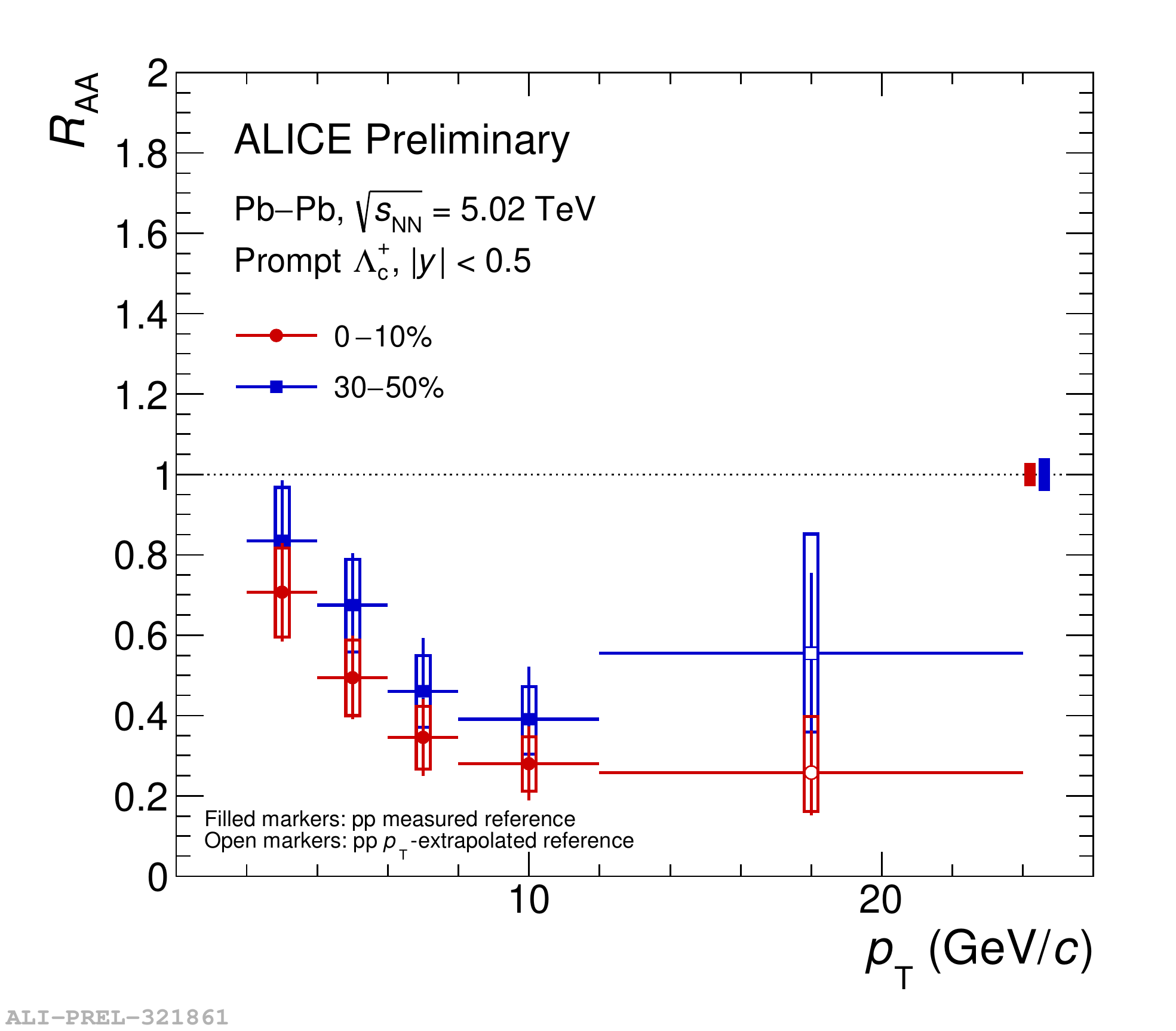}
		\hspace{0.04\textwidth}
		\includegraphics[width=.47\textwidth]{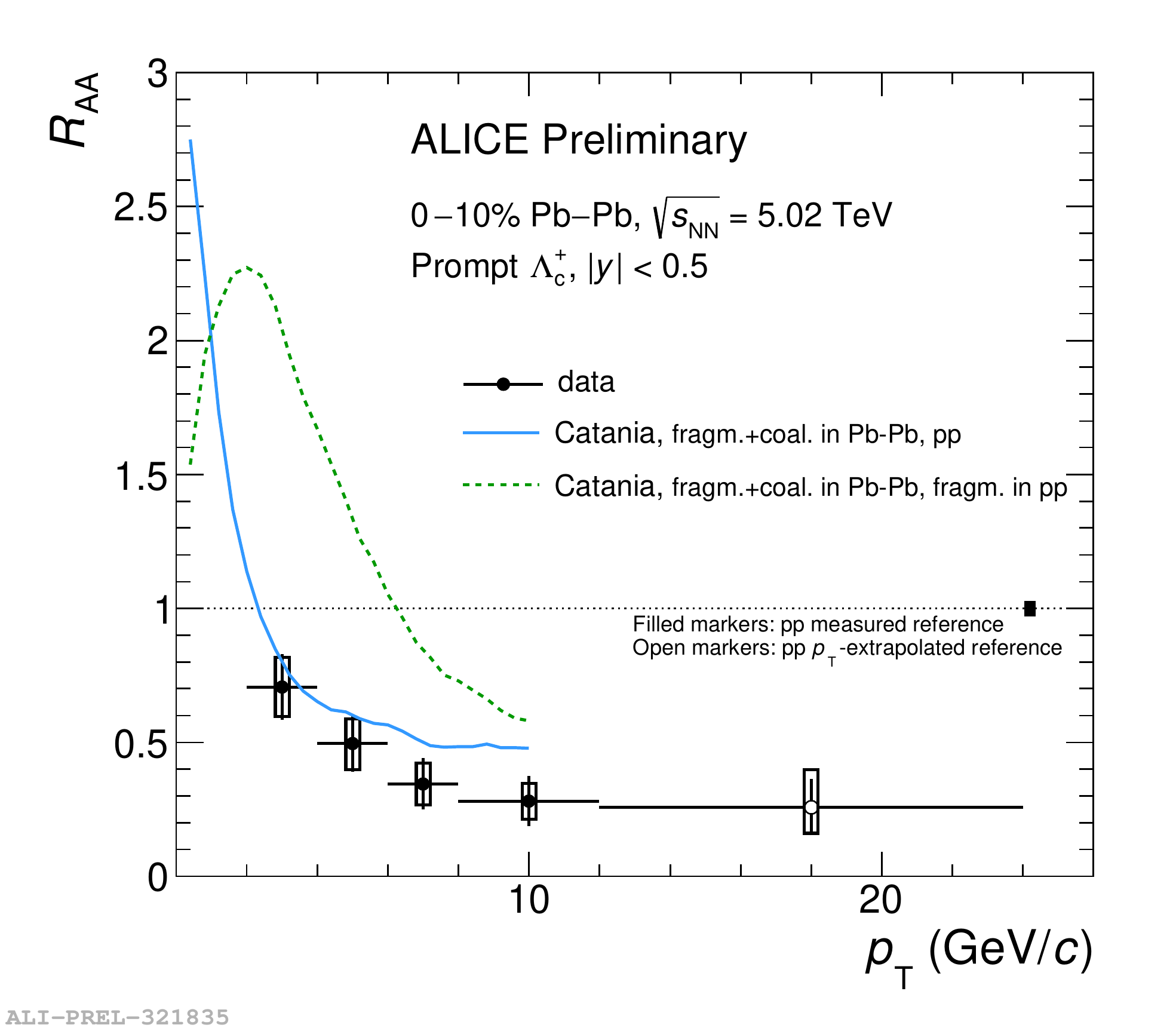}
		\caption{Left: The $R_{\rm AA}$ of the $\Lambda_{\rm c}^{+}$ baryon in the 0-10\% and 30-50\% centrality classes. Right: The $R_{\rm AA}$ of the $\Lambda_{\rm c}^{+}$ baryon in the 10\% most central collisions compared to two theoretical calculations of the Catania model~\cite{Plumari:2017ntm}.}
		\label{fig:RaaLc}
	\end{center}
\end{figure}

The left panel of Figure~\ref{fig:RaaLc} reports the results on the $R_{\rm AA}$ of the $\Lambda_{\rm c}^{+}$ baryon for $2 < p_{\rm T} < 24$~GeV$/c$ in central (0-10\%) and semi-central (30-50\%) Pb--Pb collisions, which significantly improves those based on the 2015 Pb--Pb data sample~\cite{Acharya:2018ckj}. Similar as for the ${\rm D}$ mesons, a suppression is observed for the $\Lambda_{\rm c}^{+}$ in Pb--Pb collisions, with a hint of a larger suppression for central collisions up to a factor 1.5. In the right panel of Figure~\ref{fig:RaaLc}, the $R_{\rm AA}$ for the 10\% most central collisions is compared to two theoretical calculations of the Catania model~\cite{Plumari:2017ntm}, providing two different treatments of hadronisation. The comparison favours a scenario where fragmentation and coalescence are present in both Pb--Pb and pp collisions, indicating how crucial it is to also have a better understanding of the $\Lambda_{\rm c}^{+}$-production mechanisms in pp collisions.

\begin{figure}[!tb]
	\begin{center}
		\includegraphics[width=.47\textwidth]{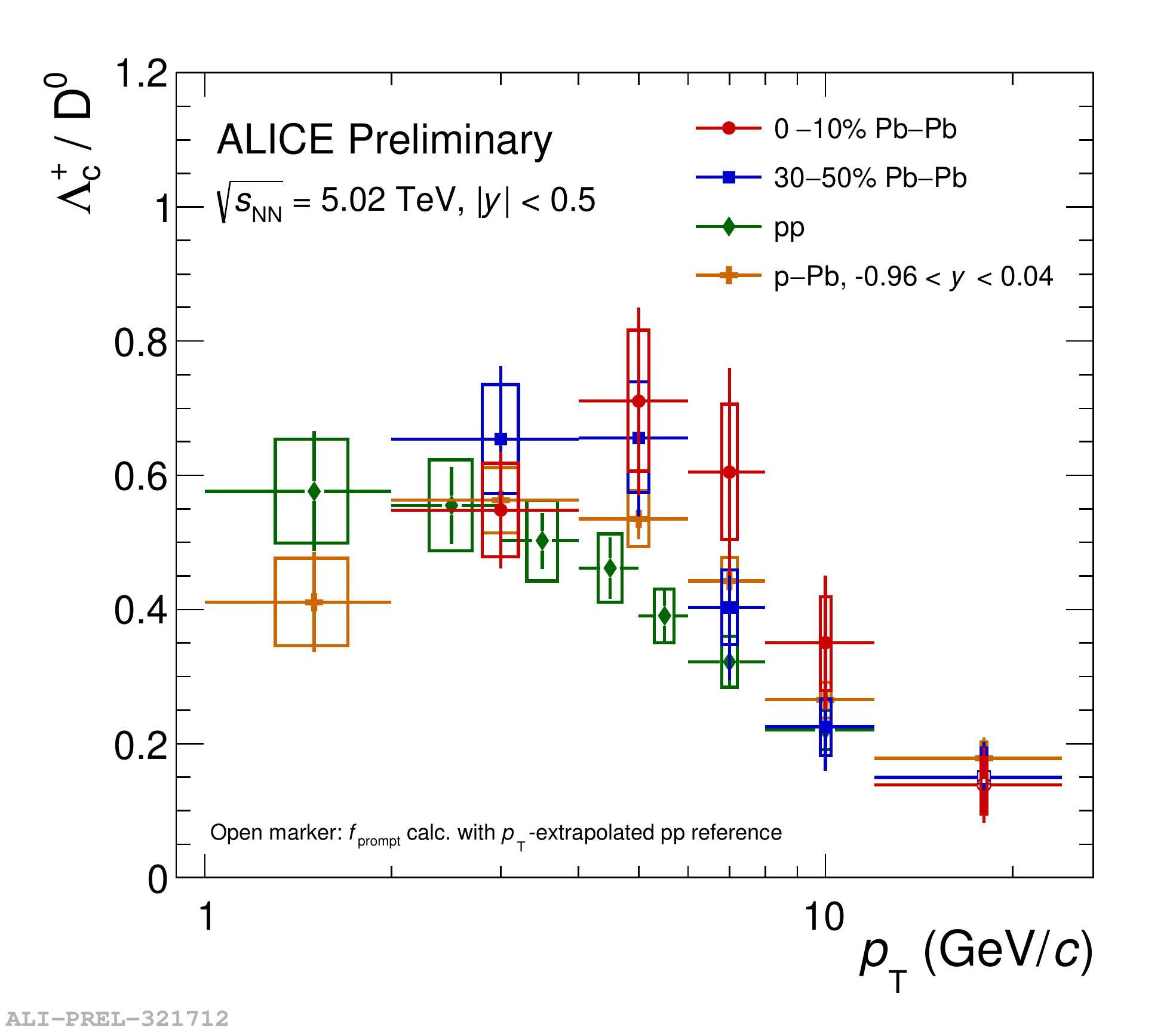}
		\hspace{0.04\textwidth}
		\includegraphics[width=.47\textwidth]{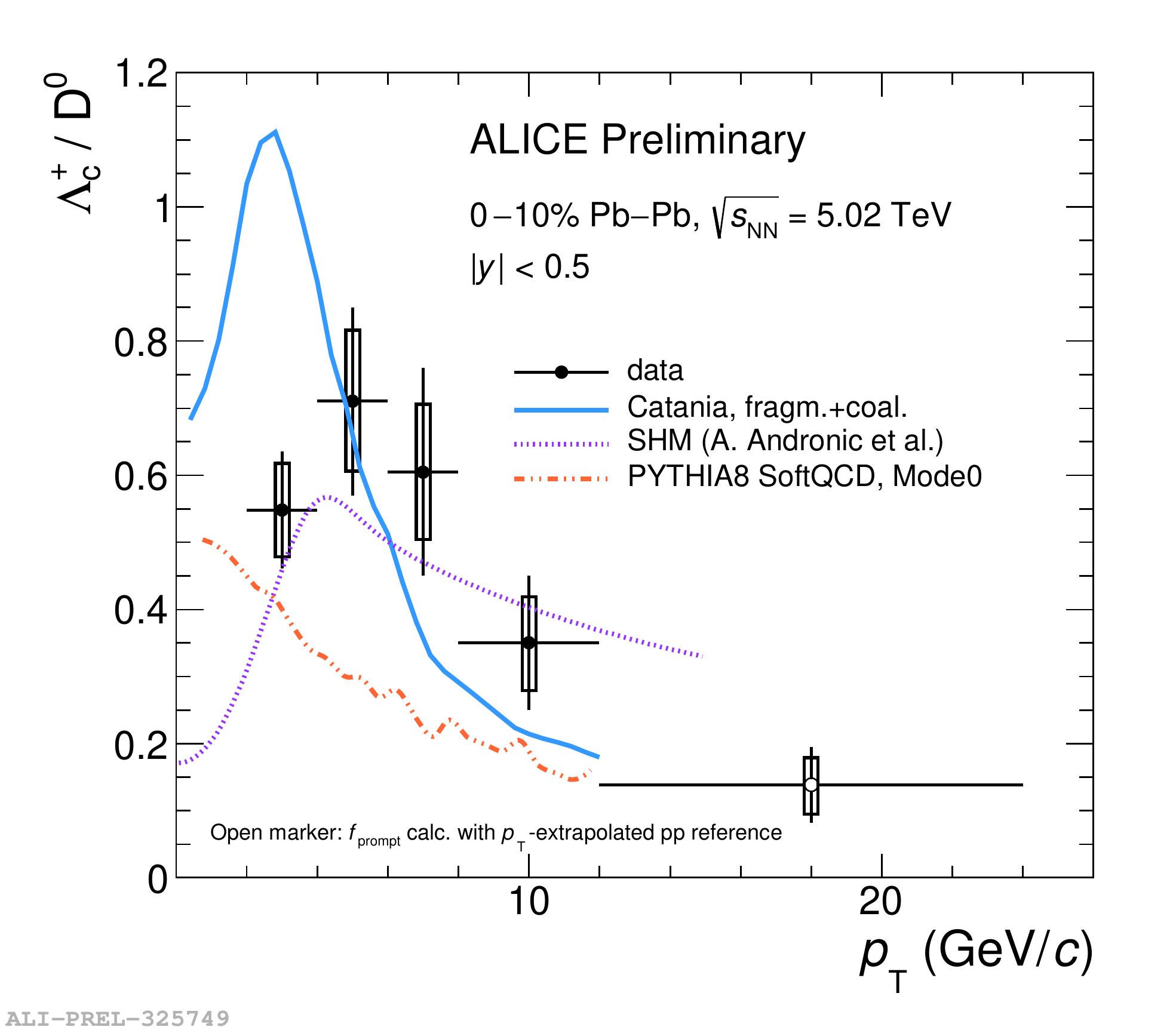}
		\caption{Left: The $\Lambda_{\rm c}^{+}/{\rm D}^{0}$ baryon-to-meson ratio in pp, p-Pb, and Pb--Pb collisions. Right: The $\Lambda_{\rm c}^{+}/{\rm D}^{0}$ baryon-to-meson ratio in the 10\% most central collisions compared with the Catania~\cite{Plumari:2017ntm}, statistical hadronisation model~\cite{Andronic:2019wva}, and PYTHIA8 with colour reconnection~\cite{Christiansen:2015yqa} model calculations.}
		\label{fig:LcoverD0}
	\end{center}
\end{figure}

In Figure~\ref{fig:LcoverD0}, the baryon-to-meson $\Lambda_{\rm c}^{+}/{\rm D}^{0}$ ratio in Pb--Pb collisions is compared to the one obtained in pp and p--Pb collisions (left panel) and to different theoretical predictions (right panel). There is a hint of a higher $\Lambda_{\rm c}^{+}/{\rm D}^{0}$ ratio in Pb--Pb with respect to pp collisions at intermediate $p_{\rm T}$ (for both the 0-10\% and 30-50\% centrality classes). More precision is needed to investigate a trend between pp, p--Pb, and Pb--Pb collisions. The measured ratio in the 0-10\% centrality class is compared to the Catania model~\cite{Plumari:2017ntm}, the statistical hadronisation model~\cite{Andronic:2019wva}, and PYTHIA8 simulations with colour reconnection~\cite{Christiansen:2015yqa}. Both the Catania and statistical hadronisation model are qualitatively describing the data, despite the different implemented hadronisation procedures. The prediction by PYTHIA8, which is intended for pp collisions, underestimates the data. Also for this observable it is fundamental to improve the understanding of $\Lambda_{\rm c}^{+}$ production in pp collisions, where many models are significantly underestimating the measured $\Lambda_{\rm c}^{+}/{\rm D}^{0}$ ratio (see Ref.~\cite{Acharya:2017kfy} for more details).

\section{Conclusions}

In this contribution, the most recent results on the production of ${\rm D}_{\rm s}^{+}$ mesons and $\Lambda_{\rm c}^{+}$ baryons in Pb--Pb collisions at $\sqrt{s_{\rm NN}}=5.02$~TeV were presented. These new measurements were performed on the latest Pb--Pb collision data sample, collected in 2018. With respect to earlier ALICE publications~\cite{Acharya:2018hre, Acharya:2018ckj}, the measurements are more differential in $p_{\rm T}$ and centrality and have an improved statistical precision. The results show hints of coalesence being an important hadronisation mechanism for heavy-flavour particles in Pb--Pb collisions and triggered again the need of a better understanding of charmed baryon production in pp collisions.

\end{document}